\documentclass{mpe_report}

\usepackage{psfig,graphicx,epsfig}
\usepackage{color}
\usepackage{amsmath,amssymb,epic,eepic,array}

\begin{document}

\pagenumbering{arabic}
\setcounter{page}{72}

\renewcommand{\FirstPageOfPaper }{ 72}\renewcommand{\LastPageOfPaper }{ 75}

\title{Giant Pulses of Pulsar Radio Emission}
\author{A.~D. Kuzmin}
\institute{Pushchino Radio Astronomy Observatory,
              Lebedev Physical Institute,
              Russian Academy of Sciences, Pushchino, 142290, Russia\\
	      e-mail: akuzmin@prao.ru}
\maketitle

\begin{abstract}
Review report of giant pulses of pulsar radio emission, 
based on our detections of four new pulsars with giant pulses, and 
the comparative analysis of the previously known pulsars with 
giant pulses, including the Crab pulsar and millisecond pulsar PSR 
B1937+21.
\end{abstract}

\section{Introduction}

Giant pulses (GPs)- a short duration outbursts- are a special form
of pulsar radio emission.

This is a most striking phenomenon of the pulsar radio emission.
For normal pulsars, the intensity of single pulses varies by no
more then one order of magnitude (Hesse \& Wielebinski 1974,
Ritchings 1976).  GPs peak flux densities can exceed hundreds and
thousands of times the peak flux density of regular pulses. GPs
are the brightest sources of the radio emission  observed in the
Universe.

This  rare phenomenon has been detected only in 11 pulsars among
more than 1\,500 known ones.
 The firsts two of them to be discovered were Crab pulsar
  PSR B0531+21  (Staelin \& Reifenstein 1968; Argyle \& Gower 1972;
  Gower \& Argyle 1972) and the millisecond pulsar PSR B1937+21
  (Wolsczcan, Cordes \& Stinebring 1984), and for a long time
  remain little investigated.

For over 20 years only these two pulsars were known to emit GPs. 
The last 5 years were marked by a systematic search of GPs that
benefit a fast progress in the detection of nine new pulsar with
GPs. They are PSR B1821-24 (Romani \& Johnston 2001), PSR B1112+50
(Ershov \& Kuzmin 2003), PSR B0540-69 (Johnston \& Romani 2003),
PSR B0031-07 (Kuzmin, Ershov \& Losovsky 2004), PSR J0218+42
(Joshi et al. 2004), PSR B1957+20 (Joshi et al. 2004), PSR
J1752+2359 (Ershov \& Kuzmin 2005), PSR J1823-3021A (Knight,
Bailes et al. 2005) and PSR B0656+14 (Kuzmin \& Ershov 2006).

\section{Observations }
GPs are observed as a rare very strong pulses of pulsar radio
emission standing out of the noise background and underlying
ordinary individual pulses. An example of one GP of the Crab
pulsar observed at frequency 111 \,MHz inside of 150 pulsar
periods is demonstrated in Fig.1.
\begin{figure}
   \resizebox{\hsize}{!}{\includegraphics{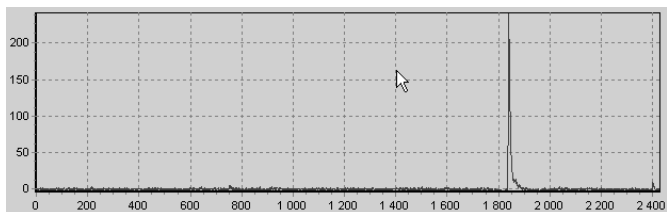}}
      \caption{ An example of a GP of the Crab pulsar inside o
      f 150 normal pulses masked by noise}
      \label{Fig1}
   \end{figure}

GPs are distinguished by a number of characteristic features.
They possess very large excess of the flux density and energy of
radio emission relative to an average pulse (AP), the energy
distribution of GPs has a  power-law statistic, have a short pulse
time-scale and occur in a narrow-phase window of an AP (except of
the Crab pulsar for which GPs can occur anywhere within the AP).

\subsection{Flux density, energy, spectra and intensity distribution.}
GPs are the most extremal phenomena of pulsar radio emission.
Their peak flux densities can exceed hundreds and thousands of
times the peak flux density of an AP. An example of a GP in comparison with an
AP of the millisecond pulsar PSR B1937+21 is shown in Fig.2. The
peak flux densitiy of GPs in this pulsar at 1650\,MHz ranges up to
65 \,kJy and exceeds the peak flux density of an AP up to factor
of $3 \times 10^5$ (Soglasnov et al. 2004). An energy of this GP
$E = S_{peak} \times \tau $, where $\tau$ is the pulse duration,
exceeds the energy of the AP by a factor of 60.

\begin{figure}
\centerline{\psfig{file=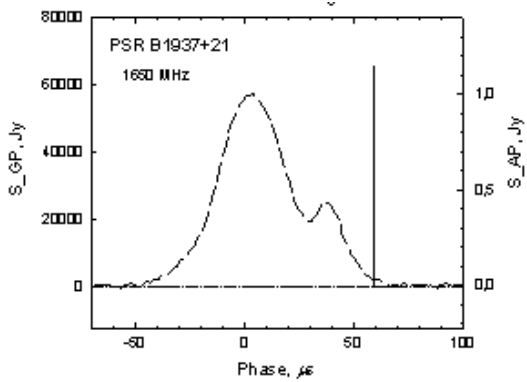,width=7.0cm,clip=} }
\caption{ An example of a GP (solid line) in comparison 
with an AP (dotted line) of the pulsar PSR B1937+21. \it (After Soglasnov et al. 2005). 
\rm Note different y-scales for a GP (left) and an AP (right).
\label{Fig2}}
\end{figure}
The peak flux densities of GPs from the Crab Nebula pulsar PSR
B0531+21 at 2228\,MHz exceeds the peak flux density of an AP  up
to factor of $5 \times 10^5$. An energy $E = S_{peak} \times \tau
$ of the GPs exceeds the energy of the AP by a factor of 80
(Kostyuk et al., 2003).

GPs of the Crab pulsar have been detected in a very wide frequency 
range from 23 \,MHz (Popov et al. 2006) up to 15 GHz (Hankins et al. 2000). 
Only small part of GPs occurs simultaneously in different frequency 
ranges. Radio spectra of these GPs were studied by simultaneous 
multi-frequency observations. 
Sallmen et al. (1999) observations of GPs in pulsar PSR
B0531+21 at two frequencies 1.4 and 0.6 \,GHz show that the GPs
spectral indices fall between -2.2 and -4.9, which may be compared
to the AP value for this pulsar -3.0. Popov et al.(2006)
observations of this pulsar at three frequencies 600, 111 and 43
\,MHz reveal that the GPs spectral indices fall between -1.6 and
-3.1 with mean value -2.7, that also may be compared to the AP
value for this pulsar. Simultaneous two-frequency observations of
GPs from PSR B1937+21 at 2210-2250 and 1414-1446 MHz (Popov \&
Stappers 2003) don't reveal any GPs which occur simultaneously in
both frequency ranges. They conclude that radio spectra of
detected GPs are limited in frequency at a scale of about $\Delta
\nu /\nu < 0.5$. Simultaneous two-frequency observations of GPs in
PSR B0031-07 at 111 and 40 MHz reveal only a few common GPs
(Kuzmin et al. 05).

Kinkhabwala et al. (2000) have estimated the average spectral
properties of the GPs emission of the pulsar PSR B1937+21. Using
the top eight GPs at three frequencies 430, 1420 and 2380 \,MHz,
they find a somewhat steeper slope of -3.1 for the GPs spectrum,
compared the -2.6 slope for the normal emission spectrum of this
pulsar.

Another distinguishing characteristic of pulsars with GPs, as
demonstrated in Fig.3, is their two-mode pulse intensity
distribution. At low intensities of ordinary pulses, the pulse
strength distribution is Gaussian one, but above a certain
threshold the pulse strength of GPs is roughly power-law
distributed (Argyle \& Gower 1972, Cognard et al. 1996) .

\begin{figure}
\centerline{\psfig{file=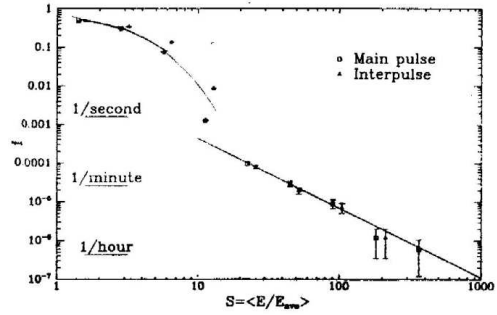,width=6.5cm,clip=} }
\caption{\textbf{a)} Cumulative distribution of GPs intensity of pulsar 
PSR B1937+21 measured in units of the mean pulse and interpulse (after Cognars et al. 1996)
\label{Fig3}}
\end{figure}
\subsection{Size and brightness temperature of an emitting region.}

The next peculiarity of GPs, as demonstrated in Fig.2 for PSR
B1937+21, is a much shorter time-scale of GP as compared with an
AP. The width of  this GP is $\tau \le 15$ \,ns (Soglasnov et al.
2004).

Hankins et al. (2003) found in pulsar PSR B0531+21 the pulse
structure as short as 2 \,ns. If one interprets this pulse
duration in terms of the maximum possible size of emitting region
$l \le c \times \tau$, where $c$ is the speed of a light, the
time-scale
$\tau=$2 \,ns corresponds to a light-travel size of an
emitting body $l$ of only 60 \,cm, the smallest object ever
detected outside our solar system.

The brightness temperature of the GPs is
$$
  T_{\rm B} = S_{peak} \lambda^2 /2k\Omega~,
$$
where $S_{peak}$ is the peak flux density, $\lambda$ is the radio
wavelength, $k$ is the Boltzmann's constant, and  $\Omega \simeq
(l/d)^2$ is the solid angle of the radio emission region.

For observed in Crab pulsar at frequency 5.5\,GHz peak flux
density of $10^3$\,Jy  with $\tau = 2$\,ns (Hankins et al. 2003)
the brightness temperature of GPs is as high as $10^{37}$ \,K.

Soglasnov et al. (2004) have proved that the majority of GPs from
the millisecond pulsar PSR B1937+21 are shorter than 15 \,ns. At a
frequency of 1.65 \,GHz a peak flux density of $65 \times 10^3$
\,Jy with $\tau = 15$ \,ns show that the brightness temperature of
GPs in this pulsar is $T_B \ge 5 \times 10^{39}$ \,K, the
brightest radio emission observed in Universe.

However, these evaluations of  size  and brightness temperature of
an emitting body are not unambiguous.  Gil \& Melikidze (2004)
argued  that the apparent duration of the observed impulse
$\tau_{obs}$ may be shorter $^{1)}$ than the duration of the
emitted one as $\tau_{rad}$ as $\tau_{obs} = \tau_{rad} \times
\gamma^{-2}$. For Lorentz factor $\gamma \approx 100$, the
time-scale of pulse structure for B1937+21 will transformed from
observed $\tau_{obs}$= 2 ns to emitted $\tau_{rad}$ =20 $\mu s$.
\footnote{Relativistic shortening of a pulse duration were claimed
by Smith (1970) and Zheleznyakov (1971).} As well as $T_B \propto
\Omega^{-1} \propto \tau^{-2}$, the brightness temperature will be
reduced to $T_B \approx 10^{31}$ \,K.

This aspect needs a further refinement.

The duration of GPs may be masked by the scatter broadening of a
pulse due to multi-path propagation in the interstellar plasma.

GPs are clustered around a small phase window (except of Crab
pulsar, for which GPs can occur anywhere within the AP). An
example of such clustering for pulsar PSR B0031-07 is shown in the
bottom of Fig.4.

\begin{figure}
\centerline{\psfig{file=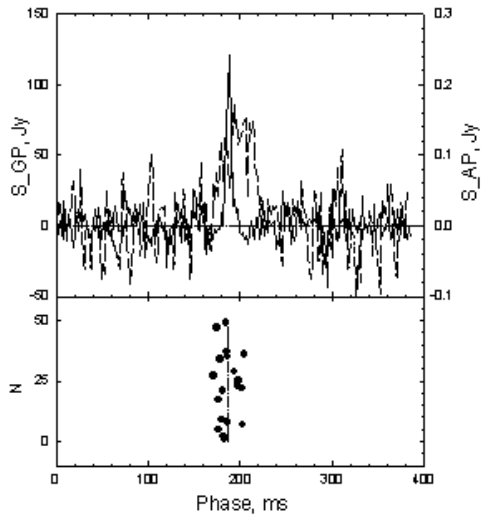,width=6.5cm,clip=} }
\caption{\textbf{(top)} GP in pulsar PSR B0656+14 at 111  MHz (bold line)
       together with the AP (dotted line). The plot of AP is presented on a
       500 times larger scale and flux densities of GP and AP are shown
       separately on the left and right sides of the "y" scales. \textbf{(bottom)} The
       phases of the observed GPs relative to the center of the
       AP. N is the number of GP.
       (after Kuzmin \& Ershov ,2006)
       \label{Fig4}}
\end{figure}

The top of Fig4 shows the GP together with the AP for the pulsar
PSR B0656+14 (Kuzmin \& Ershov 2006). The peak flux density of this 
GP exceeds the peak flux density of AP by factor of 630. 

Kuzmin \& Ershov (2004) two frequency observations of GPs in the
pulsar PSR B0031-07 reveal that some GPs have two components and
that the spacing of these components at 40 \,MHz is larger than at
111 \,MHz. This is similar to a larger width of the AP at lower
frequency, which is interpreted as a divergence of magnetic field
lines in the radius-to-frequency mapping in the hollow cone model
of pulsar radio emission. This suggests that  GPs from this pulsar
originate in the polar region rather than in the light cylinder
region.

One may suggests that there are two classes of GPs, one associated
with high-energy emission from the outer gap, the other associated
with the polar radio emission. GPs of PSR B0031--07 will be of the
second class.

\rm Table 1 summarizes the comparative data for known pulsars with
GPs. Here PSR is a pulsar name, \it P \rm is the pulsar period,
$B_{LC}$ is a magnetic field strength at the light cylinder, \it
Freq \rm is the observations frequency, $S_{GP}/S_{AP}$ is an
excess of the peak flux density of a strongest GP over the peak
flux density of an AP, $T_B$ is a brightness temperature of the
strongest GP radio emission and $E_{GP}/E_{GP}$ is an excess of
the energy of the strongest GP over the energy of an AP.

\begin{table*}[t]
  \caption[]{ Properties of the Giant Pulses}
  \label{Tab:publ-works}
  \begin{center}\begin{tabular}{clclclclc}
  \hline\noalign{\smallskip}
PSR&\it P \rm&$log B_{LC}$&\it Freq \rm& $S_{GP}/S_{AP}$ &$T_B$& $E_{GP}/E_{GP}$&References\\
    &ms &G & GHz  &             &  K   &  &  \\
      \hline\noalign{\smallskip}
 B0031-07&943&6.9&0.04&  400        &$\ge10^{28}$             & 75&Kuzmin \& Ershov 2004 \\
         & & &0.11&  120        &$\ge10^{26}$             &  8    &Kuzmin et al. 2004 \\
 J0218+42&2.3&$3.2\times 10^5$&0.61     &                 &1.3& 51&Joshi et al. 2004 \\
 B0531+21&33.4&$9.8\times 10^5$&0.14     & 300               &&   &Argyle \& Gower 1972\\
         & & &0.59&  $6\times 10^4$&$\ge10^{36}$   &  75          &Kostyuk et al. 2003 \\
         & & &2.23&  $5\times 10^5$&$\ge10^{34}$  &  80 &          Kostyuk et al. 2003 \\
         & & &5.50&                &$\ge10^{37}$  &        &       Hankins et al. 2003 \\
 B0540-69&50.5&$3.5\times 10^5$&1.38        &$5\times 10^3$& & &   Johnston \& Romani 2003 \\
 B0656+14&385&770 &0.11&  600   &$ \ge10^{26}$     & 110&          Kuzmin \& Ershov 2006 \\
 B1112+50&1656&4.1 &0.11&   80      &$ \ge10^{26}$      & 10&      Ershov \& Kuzmin 2003 \\
 J1752+2359&409&71 &0.11&  260      &$ \ge10^{28}$      &200&      Ershov \& Kuzmin 2005 \\
 B1821-24&3.0&$7.2\times 10^5$ &1.51&           &         & 81&    Romani \& Johnston 2001 \\
 J1823-3021A&5.4&$2.5\times 10^5$&0.68& 680  &             & 64&   Knight et al. 2005 \\
         & & &1.40&1700 &      &      28  &                        Knight et al. 2005 \\
 B1937+21&1.5&$9.8\times 10^5$& 0.11& 600&$\ge 10^{35}$  & 65&     Kuzmin \& Losovsky \\
         & & &1.65&  &$\ge 5\times 10^{39}$ &    60          &     Soglasnov et al. 2004 \\
 B1957+20&1.6&$3.8\times 10^5$ &0.61      &     &         &129&    Joshi et al. 2004\\
  \noalign{\smallskip}\hline
  \end{tabular}\end{center}
\end{table*}

\section{Are giant pulses inherent to some special property of pulsars ?}
The first detected pulsars with GPs PSR B0531+21 and PSR B1937+21
were among a small group of pulsars with highest magnetic field at
the light cylinder $B_{LC}=10^4 -10^5$\,G. This gave a rise to the
suggestion that GPs are inherent to pulsars with very strong
magnetic field at the light cylinder and the search of GPs was
oriented on those pulsars. As a result GPs were detected in five
other pulsars with very strong magnetic field at the light
cylinder: PSR B1821-24 (Romani \& Johnston 2001), PSR B0540-69
(Johnston \& Romani 2003), PSR J0218+42 (Joshi et al. 2004), PSR
B1957+20 (Joshi et al. 2004) and PSR J1823-3021 (Knight et al.
2005).

However detection of the GPs in four pulsar with ordinary magnetic
field at the light cylinder: PSR B0031-07 (Kuzmin, Ershov \&
Losovsky 2004), PSR B0656+14 (Kuzmin \& Ershov 2006), PSR
B1112+50 (Ershov \& Kuzmin 2003) and PSR J1752+2359 (Ershov \&
Kuzmin 2005) have revealed that GPs can occur in ordinary pulsars
too.

Johnson \& Romany (2004) claimed that GPs may be associated with
pulsars which possess a high energy X-ray emission. But the
alignment of X-ray pulses with GPs is observed only for four
objects among eleven known pulsars with GPs and is not a proper
indication for GPs.

Knight et all. (2005) argued that GPs may be indicated by large
rotation loss luminosity $\dot E \propto P^{-3} \dot P$. But in
fact the rotation loss luminosity of the known 11 pulsars with GPs
are differ by 6 orders of magnitude and  is not a proper
indication for GPs also.

\rm GPs occur in various types of pulsars in a wide range of
periods $P=1.5-1600$ \,ms and magnetic field at the light cylinder
$\log B_{LC}=4 -10^6$\,G over a wide range of radio frequencies.

\section{Summary}
\label{sect:summary}

Giant pulses  is a special  form of  pulsar radio emission, that
is characterized by very large excess of flux density and energy
of radio emission relative to an  average pulse, the power-law
statistic of the energy distribution , giant pulses occur in a
narrow-phase window of an average pulse and have a short pulse
time-scale.

The flux density of giant pulses increases over the flux density
of an average pulse $S_{GP}/S_{AP}$ up to  $5 \times 10^5$ .

Giant pulses energy excess over the energy of an average pulse is
$E_{GP}/E_{AP}$ = 50 - 200 and is nearly the same for different
magnetic field at the light cylinder, pulsar periods and
frequencies.

A light-travel size of an emitting body indicate the smallest
object ever detected outside our solar system.

Giant pulses are the brightest sources of radio emission  among
the known astronomical objects.

Giant pulses exist in various types of pulsars in a wide range of
periods , magnetic field at the light cylinder  and broad
frequency range.

\begin{acknowledgements}
This work was supported in part be the Russian Foundation of Basic
Research, Project N 05-02-16415.
\end{acknowledgements}


            \clearpage


\begin{thebibliography}{}

    \bibitem[1972]{argy72}
    Argyle, E.,\& Gower, F.R., 1972, ApJ, 175, L89

   \bibitem[1996]{cogn96}
    Cognard, I., Shrauner, J.A., Taylor, J.H.,\& Thorset, S.H.,
   1996 ApJ. 457, L81.

  \bibitem[2003]{ersh04} Ershov, A. A.,\& Kuzmin, A. D. 2003, Pis'ma v AZh, 29,
   111 (Astr. Lett., 29, 91)

   \bibitem[2005]{ersh05}
     Ershov, A.A.,\& Kuzmin, A.D. 2005,  A\&A, 443, 593

   \bibitem[2004]{gil04} Gil, J.,\& Melikidze, G. I. 2004, In: F. Camilo,
   B. M. Gaensler, eds., Young Neutron Stars and Their Environments, IAU
   Symposium 218, San Francisco: ASP, p.~321

   \bibitem[1972]{gowe72}
   Gower, F.R. \& Argyle, E.  1972, ApJ, 1715, L23

   \bibitem[2000]{hank00}
   Hankins, T.H., Wex, N.\& Wielebinski, R. 2000, in Pulsar Astronomy-2000
   and Beyong, ed. M. Kramer, IAU Colloquium 177, San Francisco: ASP, 202, p.~165

  \bibitem[2003]{hank03} Hankins, T.H.,\& Kern, J.S., Weatherall,
  J.C.,\& Eilek, J.A. 2003, Nature, 422, 141

  \bibitem[1974]{hess74}
     Hesse, T.H.,\& Wielebinski, R., 1974, A\&A, 31, 409

    \bibitem[2003]{john03} Johnston, S.,\& Romani, R. W. 2003, ApJ, 590, L95

   \bibitem[2004]{john04} Johnston, S.\& Romani, R.W. 2004,
   in Young Neutron Stars and Their Environments, IAU Symposium 218,
   ed F. Camilo, B. M. Gaensler, San Francisco: ASP, p.~315

  \bibitem[2004]{josh04} Joshi, B. C., Kramer, M., Lyne, A.G., McLaughlin, M.,
   \& Stairs I.H. 2004,
   in Young Neutron Stars and Their Environments, IAU Symposium 218,
   ed F. Camilo, B. M. Gaensler, San Francisco: ASP, p.~319

    \bibitem[2005]{kkink00} Kinkhabwala, A.,\& Thorset S.E. 2000, ApJ, 535,365

  \bibitem[2005]{knig05} Knight, H. S., Bailes, M., Manchester, R. N.,\& Ord S. M.
   2005, ApJ, 625, 951

  \bibitem[2003]{kost03}  Kostyuk, S.V., Kondratiev, V.I., Kuzmin, A.D., Popov, M.V.,
      \& Soglasnov V.A. 2003, Pis'ma v AZh, 29, 440 (Astr. Lett., 29, 387)

   \bibitem[2002]{kuzm02} Kuzmin, A.D.,\& Losovsky, B.Ya. 2002,
   Pis'ma v AZh, 28, 25 (Astr. Lett., 28, 21)

  \bibitem[2004]{kuzn04} Kuzmin, A. D., Ershov, A. A.,\& Losovsky B. Ya. 2004,
   Pis'ma v AZh, 30, 285 (Astr. Lett., 30, 247)

  \bibitem[2004]{kuzm04} Kuzmin, A. D.,\& Ershov A.A. 2004, A\&A, 427, 575

     \bibitem[2006]{kuzmin06}
     Kuzmin, A.~D. \& Ershov, A.~A.  2006, Pis'ma v
     AZh, 32, 650 (Astr. Lett., 32)

 \bibitem[2003]{popo03} Popov, M.V.,\& Stappers, B. 2003, Pis'ma v AZh, 29,
   111 (Astr. Lett., 29, 91)

   \bibitem[2006]{popov06}
     Popov, M,~V., Kuzmin, A.~D., Ulyanov, O.~M., Deshpande, A.~A., Ershov, A.~A.,
     Zakharenko, V.~V., Kondratiev, V.~I., Kostyuk, S.~V., Losovsky, B.~Ya.
    \& Soglasnov V.~A. 2006, AZh, 50, 562, (Astr. Report, 50)

    \bibitem[1976]{ritc76} Ritchings R.T. 1976, MNRAS, 176, 249

     \bibitem[2001]{romani}
     Romani, R.~W., \& Johnston, S. 2001, ApJ, 557, L93

   \bibitem[1999]{sall99}   Sallmen, S., Backer D.~C, Hankins, T.~H., Moffet. D.
   \& Lundgren. S. 1999, ApJ, 517, 460.

   \bibitem[1970]{smit70} Smith F.G. 1970, MNRAS, 149, 1

   \bibitem[2005]{sogl05}
   Soglasnov, V.A., Popov, M.V., Bartel N., Cannon, W., Novikov, A.Yu.,
   Kondratiev, V.I.,\& Altunin, V.I., 2004, ApJ, 616, 439

  \bibitem[1968]{stae68} Staelin, D. H.,\& Reifenstein E,C. 1968, Science,
162, 1481

\bibitem[1970]{stae70} Staelin, D. H.,\& Sutton J. M. 1970, Nature, 226, 69

\bibitem[1984]{wols84} Wolszczan, A., Cordes, J. M., \& Stinebring D. R. 1984, In:
 in Millisecond Pulsars, ed. S. P. Reynolds and D. R. Stinebring, NRAO,
   Green Bank, p.~63

  \bibitem[1971]{zhel71} Zheleznyakov V.V. 1971, ApSS, 13, 87


\end{thebibliography}
\end{document}